# Chiral ferronematic liquid crystals: a physico-chemical analysis of phase transitions and induced helical twisting

K. G. Nazarenko[a,b], N. A. Kasian[c], S. S. Minenko[c], O. M. Samoilov[c], V. G. Nazarenko[d], L. N. Lisetski[c*], I. A. Gvozdovskyy[e**]

*[a]Department of Organophosphorus Compounds Chemistry, Institute of Organic Chemistry of the NAS of Ukraine, Kyiv, Ukraine; [b]Enamine Ltd, Kyiv, Ukraine; [c]Department of Nanostructured Materials, Institute for Scintillation Materials of STC "Institute for Single Crystals" of the National Academy of Sciences of Ukraine, Kharkiv, Ukraine; [d]Department of Physics of Crystals, Institute of Physics of the National Academy of Sciences of Ukraine; [e]Department of Optical Quantum Electronics, Institute of Physics of the National Academy of Sciences of Ukraine;.*

Institute for Scintillation Materials of the National Academy of Sciences of Ukraine, Nauki Avenue 60, Kharkiv, 61072, Ukraine, telephone number: +380 57 341 0358, [*]E-mail: lisetski@isma.kharkov.ua

Institute of Physics of the National Academy of Sciences of Ukraine, Prospekt Nauki 46, Kyiv-28, 03028, Ukraine, telephone number: +380 44 525 0862, [**]E-mail: igvozd@gmail.com

# Chiral ferronematic liquid crystals: a physico-chemical analysis of phase transitions and induced helical twisting

A possibility was assessed to modify physico-chemical properties of ferronematic mixtures by introducing additional mesogenic and non-mesogenic components. The mixture, consisting of RM734 and DIO at the 70:30 weight ratio of the components, was doped with standard nematic 5CB and its chiral isomer CB15. This allowed a substantial lowering of the ferronematic temperature range, as well as induction of helical twisting in both nematic (N) and ferronematic ($N_F$) phases. The effects of 5CB and CB15 on the $N_F$ to N phase transition were nearly identical, but the isotropic transition temperature decreased more strongly upon addition of CB15. Optical transmission vs. temperature measurements showed lower transmittance in the $N_F$ phase, probably due to scattering on the domain boundaries. In the vicinity of the $N_F$ - N transition, a noticeable range of anomalously low transmission was noted both on heating and cooling, suggesting a complex character of molecular reorientation features. In the chiral $N_F$ ($N^*_F$) system, helical twisting was recorded using Grandjean-Cano wedge, with the helical pitch decreasing upon cooling in both $N^*$ and $N^*_F$ phases. In the cooling mode, transient undulations emerged in the $N^*_F$ phase. The complex picture of the ferronematic transition is supported by POM images taken at different temperatures.

**Introduction**

Recently, synthesis has been reported of several new mesogenic compounds which, alongside with the conventional nematic phase, can form, at lower temperatures, a novel type of liquid crystal arrangement, characterized by a certain kind of polar ordering [1-3].

Detailed studies of one of these substances 4-[(4-nitrophenoxy)carbonyl]phenyl 2,4-dimethoxybenzoate (RM734) reported in [4] allowed unambiguous definition of the low-temperature phase as "ferroelectric nematic" ($N_F$), characterized by the presence of domains inside which the molecules tend to parallel orientation of their dipole moments. Further generalizations and insights into the nature of this phase were presented in [5]. Another substance of the same type, 2,3',4',5'-tetrafluoro-[1,1'-biphenyl]-4-yl 2,6-difluoro-4-(5-propyl-1,3-dioxane-2-yl)benzoate (DIO) [3] was used in [6] to obtain binary phase diagrams of the RM734 - DIO system, exhibiting regions with nematic (N), ferroelectric nematic ($N_F$), and antiferroelectric smectic phase between the two nematic phases at high concentrations of DIO. In a further development, yet another substance was obtained, showing the N $\rightarrow$ $N_F$ phase transition at room temperatures [7], and chiral ferronematic liquid crystals ($N_F^*$, in parallel with well-known chiral nematics $N^*$) were reported in [8].

The polar ordering, *i.e.*, alignment of all the dipoles along a single direction within each ferroelectric domain (with approximately equal number of domains with opposite polarities) makes the $N_F$ phase substantially more sensitive to electric fields as compared with conventional nematics. This opens multiple new ways for applications of $N_F$ in electrooptics and many other fields. The realization of these prospects will require preparation of various liquid crystal mixtures that would form $N_F$ phase in the desired temperature ranges, as well as meet other requirements of each specific application. A standard way to obtain such mixtures is modifying the properties of a liquid crystal system by introducing various components of different chemical nature, which, while maintaining the basic properties of $N_F$ phase, would adjust its physical and physico-chemical properties in accordance with practical requirements.

An approach that is the closest to our consideration is that used to obtain the electromagnetically tunable cholesterics with oblique helicoidal structure [9]. These "oblique cholesterics" were, in fact, twisted high-temperature phases of mixtures forming twist-bend nematics ($N_{tb}$) that included additional standard nematic components to adjust the temperature range [10]. The effects of adding nematic (5CB) and/or cholesteric/chiral components (*e.g.*, cholesteryl oleyl carbonate or R811) to the initial $N_{tb}$-forming mixture were studied in detail by means of DSC, optical transmission, and optical microscopy [11].

Thus, the idea of our present study is the investigation of the phase transition and search for low temperature $N_F$ phase in the RM734-DIO mixture with adding a certain amount of standard nematic (5CB). One could expect that its relatively large molecular dipole would probably prevent quick suppression of $N_F$ phase. In this way, we endeavored obtaining RM734-DIO-5CB mixtures showing well-defined N $\rightarrow$ $N_F$ phase transitions in a temperature range easily accessible for water thermostat-controlled installations. As a further step, an induced chiral $N^*_F$ phase was expected by replacing non-chiral 5CB by its chiral analogue CB15. This would allow a comparison between $N^*_F$ phases formed by molecules with intrinsic chirality (such as described in [8]) and $N^*_F$ phases with helical twisting induced by chiral dopants.

**Experiment**

As a basic system for our studies, we used mixtures of RM734 and DIO (synthesized in Institute of Organic Chemistry of the NAS of Ukraine, Kyiv, Ukraine), with their binary phase diagrams presented in [6,12] serving as a pointer in determining their composition. Thus, we chose a mixture of 70% RM734 and 30% DIO as an initial object, providing the lowest N - $N_F$ transition temperature without the danger of encountering the intermediary antiferroelectric smectic phase (which could obscure the observed picture of N - $N_F$ transition).

To study the influence of different amounts of dopants on the N - $N_F$ phase transition, we chose the nematic 5CB (obtained from Chemical Reagents Plant, Kharkiv, Ukraine) and right-handed chiral CB15 (obtained from Licrystal, Merck, Darmstadt, Germany) with helical twisting power (HTP, $β$) of ~ 8 $μm^{-1}$. [13] These dopants were added to the 70% RM734 and 30% DIO mixture in concentrations in the range of 0 - 30 wt.%.

To obtain planar alignment of $N^*_F$, we used polyimide PI2555 (HD MicroSystems, USA), [14] possessing strong azimuthal anchoring energy. [15] The PI2555 films were deposited on glass substrates by spin-coating method (6800 rpm, 10 s) with subsequent annealing at 180 $^0C$ for 30 min. The polyimide layer was unidirectionally rubbed, and wedge-like cells were assembled using these substrates with antiparallel rubbing on both aligning surfaces.

To measure the gap in the wedge-like LC cell, we used the interference method, recording the transmission spectrum of empty cell (*e.g.* in the thin end $d_0$ and thick end *d*) by means of spectrometer (Ocean Optics 4000USB, USA). The thickness of the gap was set to 7 *μ*m by a spacer, and wedge-like cells with the wedge angle of 0.023 degrees were filled with $N^*_F$ using capillary action at temperature of the isotropic phase (Iso) of the mixture.

To determine the helical pitch $P$ in $N^*$ and $N^*_F$ phases, the Grandjean–Cano (GC) method was used [16-19]. It is known that in the wedge-like LC cell, with thickness of thin end $d_0$ and thickness of thick end $d$, the helical pitch $P$ in $N^*$ phase is related to the number $N_{GC}$ of the Grandjean–Cano lines (observed with the distance between them equal to half-pitch $P/2$) as follows: [16,17]

$$P = \frac{2 \cdot (d - d_0)}{N_{GC}} \quad (1)$$

To determine the helical pitch in $N^*_F$ phase it is necessary to take into account the fact that the $N^*_F$ phase in wedge-like LC cell forms the GC lines separated by Cano disclinations in such a way that neighboring lines differ by the full helical pitch P [18,19]. Due to this fact, for $N^*_F$ phase the Equation (1) should be rewritten as follows:

$$P = \frac{d - d_0}{N_{GC}} \quad (2)$$

To find out the twisting sense of the induced $N^*$ and $N^*_F$ phases, we used the GC method based on determination of the direction of interference colours shift in the wedge-like LC cell as described in detail elsewhere. [13,20] It is known that for the right-handed (*i.e.*, wave number $q_0 = 4\pi \times \beta \times C$ is positive, $q_0 > 0$) cholesteric helix, the interference fringes are shifted towards regions at the thin end of wedge when the polarizer (analyzer) rotates in the clockwise (or counterclockwise, depending upon observation conditions) direction, and vice versa for the left-handed ($q_0 < 0$) cholesteric helix. [20]

To asses the possibility of treating these components in a manner typical for other liquid crystal mixtures, we carried out thermogravimetric analysis (TGA) of the components (Mettler TA 3000, Switzerland). The results have shown that quick heating to the Iso leads to mass losses of less than ~1%, which allowed us to use the standard procedure of introducing additional components in the isotropic state with subsequent cell filling by capillary forces.

Differential scanning calorimetry (DSC) studies were performed using a microcalorimeter Mettler DSC 1 (Mettler Toledo, Switzerland). The mixtures (~20 mg) were placed into an aluminum crucible, sealed, and thermograms were recorded in consecutive scans on heating and cooling (scanning rate 5 °C/min). The procedure was repeated 4 - 5 times. Experimental error was ± 0.1 °C for the isotropic transition temperature and ± 0.2 °C for the transition to $N_F$ phase.

The textures of the $N^*_F$ mixtures were studied during both the cooling and heating processes using polarizing optical microscopy (POM). Samples were placed in a thermostable heater based on a temperature stage MikRa 603 (LLD 'MikRa', Kyiv, Ukraine). The temperature measurement accuracy was ± 0.1 ºC.

Optical transmission spectra were measured in sandwich-type LC cells, assembled using pair substrates with rubbed polyvinyl alcohol (PVA) films, using a Shimadzu UV-2450 spectrophotometer (Shimadzu, Japan) within spectral range 300-900 nm. The measurements were carried out within the temperature range of 20 - 90 °C both on heating and cooling, and the temperature values were changed in 0.2 – 0.5 °C steps and stabilized using a flowing-water thermostat (± 0.1 °C).

## 3. Results and discussions

### 3.1. Differential scanning calorimetry

The basic mixture of ferroelectric nematics consisting of 70% RM734 and 30% DIO was doped with the nematic 5CB and its non-mesogenic chiral isomer CB15 in different concentrations. The phase transition temperatures of the mixtures were obtained by DSC and checked using POM. The DSC results are shown in Figure 1. The obtained phase diagram is rather typical for liquid crystal systems. One can note the following features.

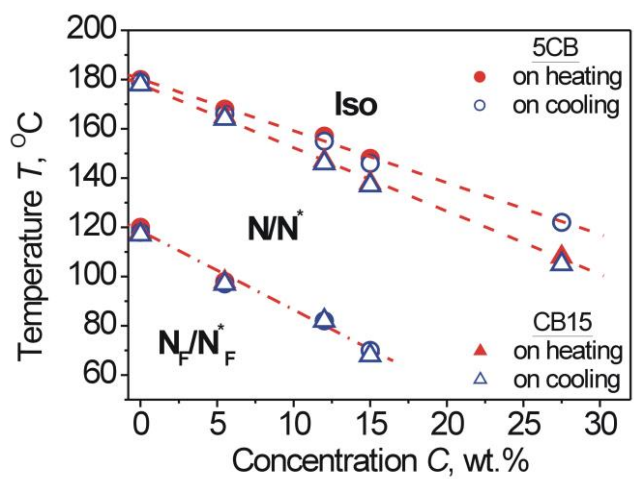

Figure 1. Phase transition temperatures between isotropic (Iso), nematic/chiral nematic (N/N*) and ferroelectric nematic/chiral nematic ($N_F/N^*_F$) phases for the basic mixture 70

wt.% RM734 and 30 wt.% DIO as a function of concentration of nematic 5CB (circle symbols) and chiral dopant CB15 (triangle symbols).

It was found that the temperatures of both phase transitions ($N/N^* \rightarrow$ Iso and $N/N^* \rightarrow N_F/N^*_F$) decrease almost linearly with dopant concentration, suggesting the absence of specific intermolecular interactions between the dopant and the matrix. The isotropic transition can be clearly determined up to dopant concentration of ~ 30 wt.%; however, DSC peaks of the ferroelectric transition cannot be reproducibly measured above ~15 wt.% due to appearance of solid phase.

It is interestingly that there is practically no difference in the effects of 5CB and CB15 upon the $N/N^* \rightarrow N_F/N^*_F$ transition. At the same time, CB15 decreases the isotropic transition temperature more markedly as compared with 5CB. This is quite natural since 5CB is a nematogenic substance, thus it deteriorates the orientational order only slightly due to is lower nematic-isotropic transition temperature being lower than that of the basic $N_F$ mixture; the effect of non-mesogenic CB15 is noticeably stronger. This difference between 5CB and CB15 is obviously irrelevant for the ferroelectric ordering, so, the effects of 5CB and CB15 upon the ferroelectric transition temperatures are almost identical. A qualitatively similar behaviour was recently noted in [21], where mixtures of two DIO isomers (*trans*- and *cis*-) were considered – addition of non-mesogenic *cis*-DIO to mesogenic *trans*-DIO strongly decreased the nematic to isotropic transition temperatures, but showed much weaker effect upon thermal stability of the ferroelectric nematic phase.

## 3.2. Polarizing optical microscopy studies of the chiral nematic to chiral ferronematic phase transition

We used a wedge-like LC cell (10 mm length, 15 mm width, ~ 2 µm and ~ 7 µm thickness at the ends) filled by the chiral mixture consisting of 15 wt.% CB15 dissolved in our basic mixture of $N_F$ compounds (*i.e.* 70 wt.% RM734 and 30 wt.% DIO). The texture images obtained at various temperatures are shown in Figure 2.

In the case of tangential boundary conditions the $N^*$ phase is characterized by the GC (or *planar*) texture (Figure 2(a)) possessing GC lines separating the neighboring areas by Cano disclinations. [16,17,28] Each neighboring GC line marks a jump in the cholesteric pitch $P$ (*i.e.*, the $P$ values in the neighboring areas differ by $(n+1/2)\pi$).

[28] The wedge-like LC cell images in $N^*_F$ phase at 62 ºC and solid quasi-crystalline ($N^*_F$ glassy) phase at 22 ºC are shown in Figure 2(b) and Figure 2(c), respectively. Figure 2 (d) shows the typical GC texture of the $N^*$ phase at 80 ºC.

In contrast to [8] no change in the GC texture was observed during slow cooling (~ 0.02 ºC/min) of the wedge-like LC cell (Figure 3), while fast cooling process (~ 5 ºC/min) to $N^*_F$ results in the appearance of light scattering in the area neigbouring GC lines, as can be seen in Figure 2 (b). The light scattering areas (Figure 2 (e)) are characterized by the emerging texture similar to undulations that can be observed for instance, in thin planar cholesteric layer under the alternating electric/magnetic field. [29-34] In Figure 2 (e) the texture with 2D undulations appears just below the temperature of $N^* \rightarrow N^*_F$ phase transition.

Under cooling of the sample to the room temperature the transition "$N^*_F$ liquid to $N^*_F$ glassy" occurs. At the beginning, this is accompanied by light scattering due to appearance of the "star" texture after 20-30 min storage, as can be seen from Figure 2 (c), (f). Further storage of sample at 22 ºC in 2 – 3 hours leads to the formation of the $N^*_F$ glassy phase (Figure 2 (g)).

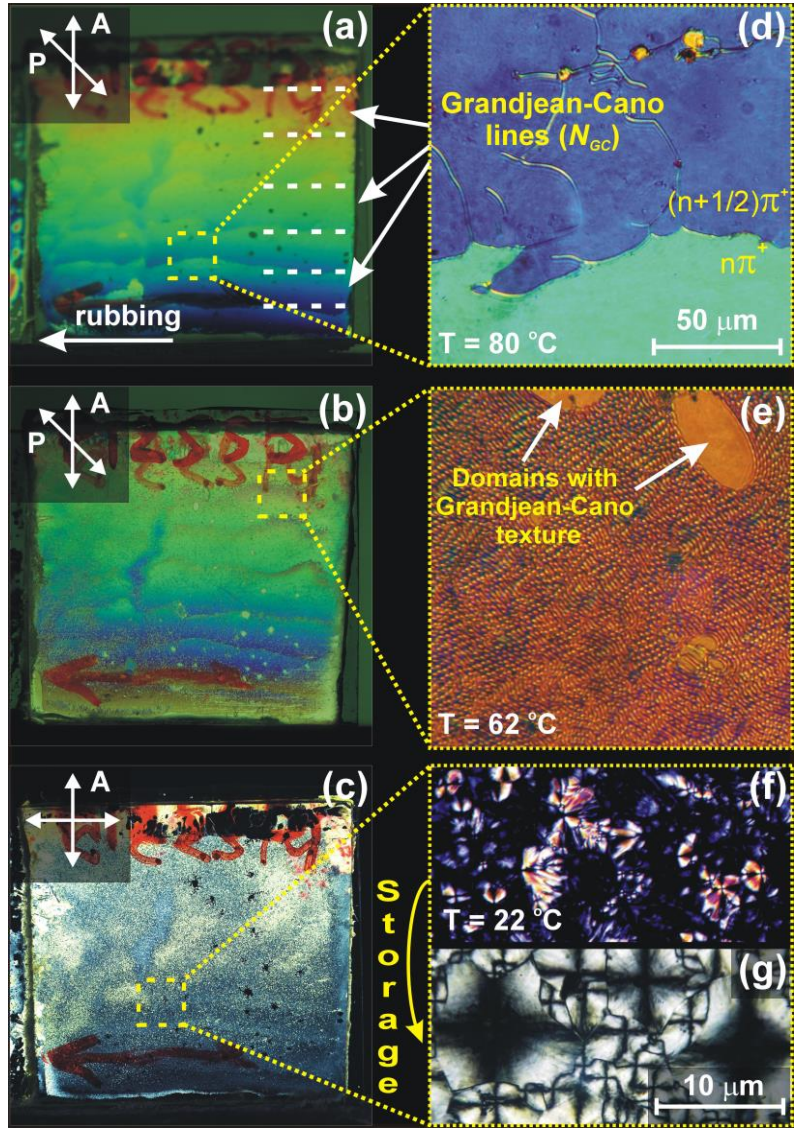

Figure 2. Photography of the wedge-like LC cell filled with N*$_F$ mixture (*i.e.* RM734:DIO (70:30) doped with 15.5 wt.% CB15) between crossed polarizers at various temperatures: (a) 80 °C – chiral nematic phase N*; (b) 62 °C – chiral ferroelectric nematic phase N*$_F$ and (c) 22 °C – N*$_F$ glassy phase. Microphotography of textures observed in POM at various temperatures: (d) 80°C – GC texture; (e) 62 °C – 2D undulation of N*$_F$ helix under temperature; (f) 22 °C – N*$_F$ glassy texture; (g) 22 °C - N*$_F$ glassy texture after 2-3 hours of storage. Wedge-like LC cell characterized by thin end $d_0 = 2.7$ µm and thick end $d = 6.8$ µm. The number of GC lines $N_{GC} = 6$ at T = 80 °C. The estimated helical pitch length $P$ is about 1.3 µm.

It should also be noted that similarly to N* phase with the helix pitch $P$ about a few microns, we observed the undulation structures formed in N*$_F$ phase. By taking into account the ratio between the thickness $d$ of LC cell and the length of N*$_F$ helix $P$, the undulation structures are formed during the phase transition $N^* \rightarrow N^*_F$ on cooling and can possess both 1D and 2D structures. For instance, at the thin end of wedge LC cell (*e.g.* $d$ = 2.7 µm), where the $d/P$ ~ 2.7, the appearance of 1D undulation structure perpendicular to the rubbing direction with the period $\Lambda$ = 5 µm was observed (Figure 3 (a)). In the thick area of the wedge-like LC cell (*e.g.* $d$ = 6.7 µm), where the ratio $d/P$ is approximately equal to ~7 or more, the 2D undulation structures appeared having period both in horizontal ($\Lambda_h$) and vertical ($\Lambda_v$) direction about 7 µm (Figure 3 (b)).

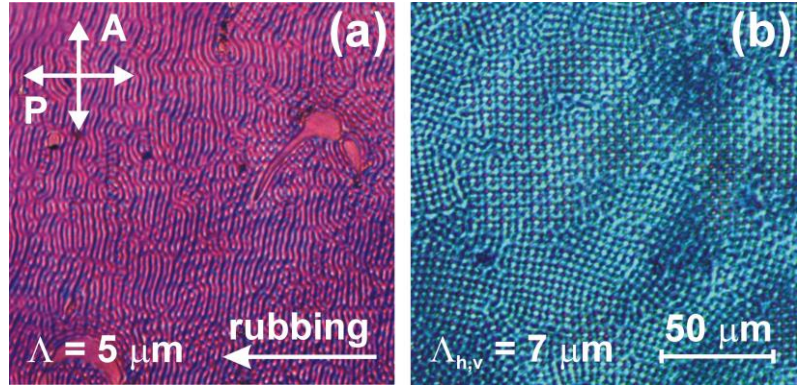

Figure 3. The $N^*_F$ undulation structures formed at: (a) thin and (b) thick end of the wedge-like LC cell during fast cooling to 42 ºC. $N^*_F$ mixture contains 15.5 wt.% of CB15 dissolved in basic $N_F$ (*i.e.* mixture of the 70 wt.% RM734 and 30 wt.% DIO). At 24 ºC the helical pitch $P$ of the $N^*_F$ was ~ 1 µm. The ratio $d/P$ was ~ 2.7 (a) and ~ 7 (b). The period of undulation structures $\Lambda$ is: (a) 5 µm and (b) 7 µm for both horizontal (h) and vertical (v) direction.

Unlike undulations in $N^*$ phase formed during alternating electric field [30-34], the undulation structures in $N^*_F$ phase appear during temperature gradient at about 3 - 5 ºC below the temperature of the $N^* \rightarrow N^*_F$ transition. The appearance of the domains is shown in Figure 4. At the beginning, the domains with GC texture separated by walls were observed (Figure 4(a)). Obviously, these domains are characterized by opposite polarization in $N^*_F$, as it was studied in detail in [8]. The fast decreasing of temperature leads to the appearance of undulation structures in domains with opposite helix handedness. Domains with GC texture are separated by walls from domains with undulation structures (Figure 4 (b)).

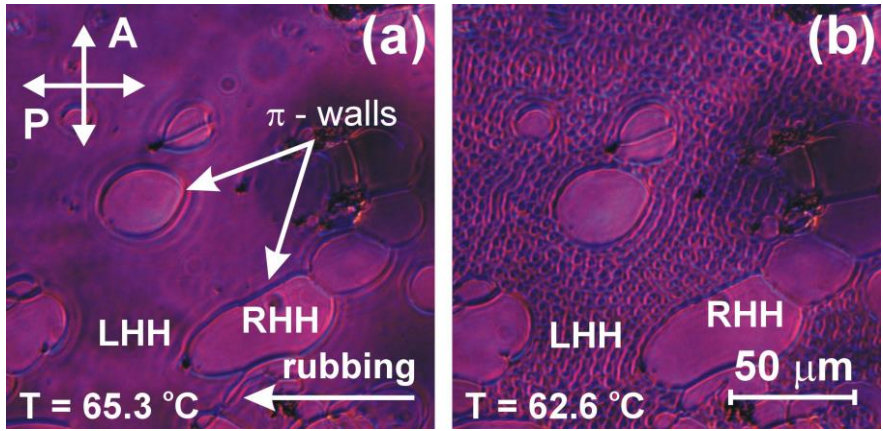

Figure 4. Photography of domains with opposite handednes of the $N^*_F$ phase, based on right-handed CB15 and basic $N_F$ mixture (70 wt.% RM734 and 30 wt.% DIO). (a) Domains with GC textures having opposite handednes $N^*_F$ phase at about 4 °C below than the temperature of the $N^* \to N^*_F$ transition. (b) GC texture for the right-handed helix (RHH) and undulation structure for the left-handed helix (LHH) $N^*_F$ phase at 62.6 °C after formation of the undulations. Thickness of wedge-like LC cell is about 5.2 μm.

The phase transition $N^* \to N^*_F$ was observed in the wedge-like LC cell with thick end $d \sim 4.5$ μm during slow cooling of the $N^*$ phase that characterized by the GC texture as shown in Figure 5 (a). As well, slow cooling process leads to the rapid flow of the LC material in $N^*$ phase at temperature close to the phase transition (Figure 5 (b),(c)). The direction of the flow was always from thin to thick end of LC cell (*i.e.* for our experimental conditions the flow is perpendicular to rubbing direction).

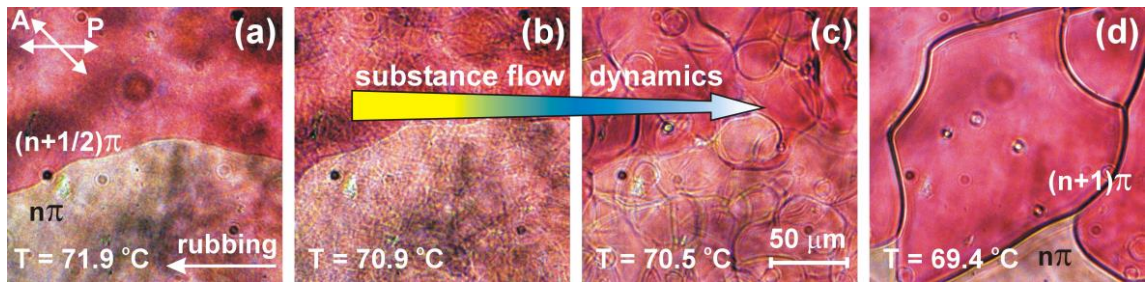

Figure 5. Phase transition on cooling from $N^*$ (a) through unstable dynamic process ((b) and (c)) to $N^*_F$ phase (d) in wedge-like LC cell filled with the mixture RM734:DIO (70:30) doped with 15.5 wt.% CB15.

As distinct from the fast cooling, when GC texture of $N^*$ phase transforms to undulations in the $N^*_F$ phase (Figure 3, Figure 4), slow cooling process of $N^*$ phase with the GC texture leads to the appearance of the $N^*_F$ phase having GC texture with distinctive domains (Figure 5(d)).

To determine the helix handedness in $N^*_F$ phase, we used a simple and accurate GC method, [20] which is often applied for determination of $N^*$ helical twisting in a wedge-like LC cell. Figure 6 shows the wedge-like LC cell, filled by basic $N_F$, $N^*$ and $N^*_F$. The LC cell was placed between crossed polarizer (P) and analyzer (A). The rotation of polarizer in the clockwise direction leads to shift of interference stripes to thin/thick end of the wedge-like LC cell for right/left-handed (RH/LH) helix. It should be noted that interference stripes are parallel to the Cano disclinations. For instance, for standard chiral nematic mixture, based on the 11.4 wt.% of CB15 dissolved in the nematic E7, the RH helix was determined.

As shown in [35], $N_F$ phase in a sandwich LC cell with substrates rubbed in anti-parallel directions is characterized by the structure showing RH twist or LH twist in different domains. In case of the basic $N_F$ mixture (*i.e.* 70 wt.% RM734 and 30 wt.% DIO), the rotation of polarizer in the clockwise direction (*i.e.* the usage of GC method) leads to the shift of stripes in opposite direction, namely to the thin end of wedge-like LC cell for RH domains and thick end for LH domains (Figure 6 (a), (b)). It can be easily seen that interference stripe (*e.g.* blue colour in Figure 6 (a)) is located at a certain thickness *d* of the wedge (it is marked by white dashed line) and there is the same behaviour for both RH and LH domains. Figure 6 (b) shows the rotation of polarizer in the clockwise direction by 30 degrees, leading to the shift of interference colours in opposite directions (for blue colour, it is marked by white arrows), which moved to the thin end of wedge $d_1$ for domains with RH helix and to the thick end $d_2$ for domains with LH helix.

It should be noted that the slow cooling of the sample, when GC texture is stored, the shift of interference colours occurs only towards thin end of the wedge-like LC cell for both $N^*$ (Figure 6 (c), (d)) and $N^*_F$ phase (Figure 6 (e), (f)) which is typical for medium having RH helix (Figure 6 (c), (d)). The shift of colours was determined near the phase transition $N^* \rightarrow N^*_F$ (*i.e.* at about 86 ºC) of the $N^*_F$ mixture containing 11.1 wt.% CB15 and the basic $N_F$ mixture. As distinct from the recently studied chiral ferronematic compound II-4$^*$ [8] for our $N^*_F$ mixture induced by means of the right-handed chiral dopant CB15 no inversion of the helix of $N^*_F$ was observed.

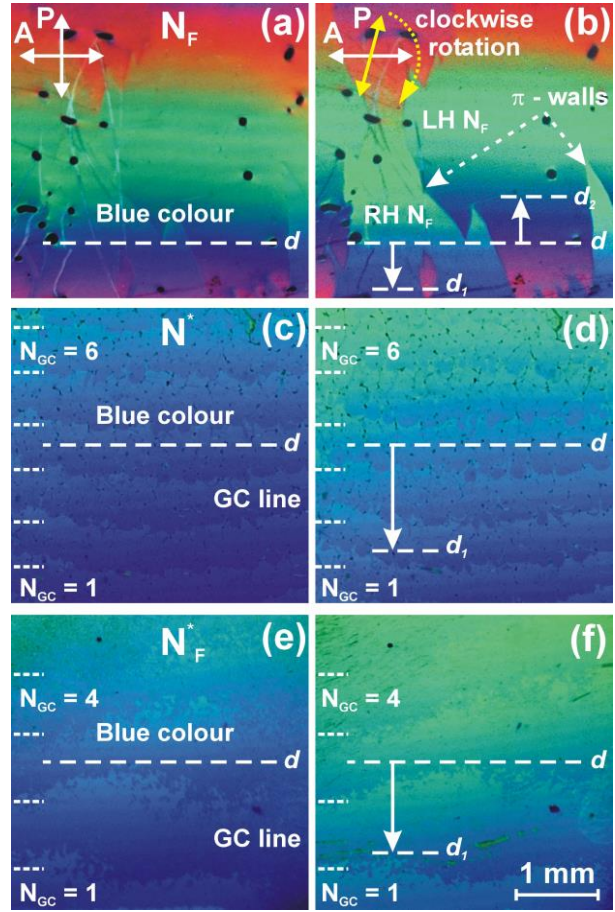

Figure 6. The shift of the interference colours in wedge-like LC cells during rotation of polarizer in the clockwise direction (GC method) for various phases: (a), (b)- basic $N_F$ mixture at 120 ºC; (c), (d) - $N^*$ phase at 89 ºC and (e), (f) - $N^*_F$ phase at 84 ºC of the chiral $N_F$ mixture RM734:DIO (70:30) doped with 11.1 wt,% CB15.

In addition we studied the dependence of the helical pitch $P$ in various phases during the cooling process of our $N^*_F$ mixture containing 11.1 wt.% CB15. The calculations of the helical pitch $P$ in $N^*$ and $N^*_F$ phases were carried out using Equation (1) and (2), respectively.

The change of $N_{GC}$ during the cooling process for the $N^*$ and $N^*_F$ phase is shown in Figure 7. It is seen that the decreasing of temperature $T$ in $N^*$ phase (Figure 7 (a)-(c)) increases the number of GC lines ($N_{GC}$). By using the geometrical dimensions of the wedge-like LC cell (Figure 7(b)) and Equation (1) the calculated helical pitch $P$ values are shown in Figure 8 (opened red circles, curve 1).

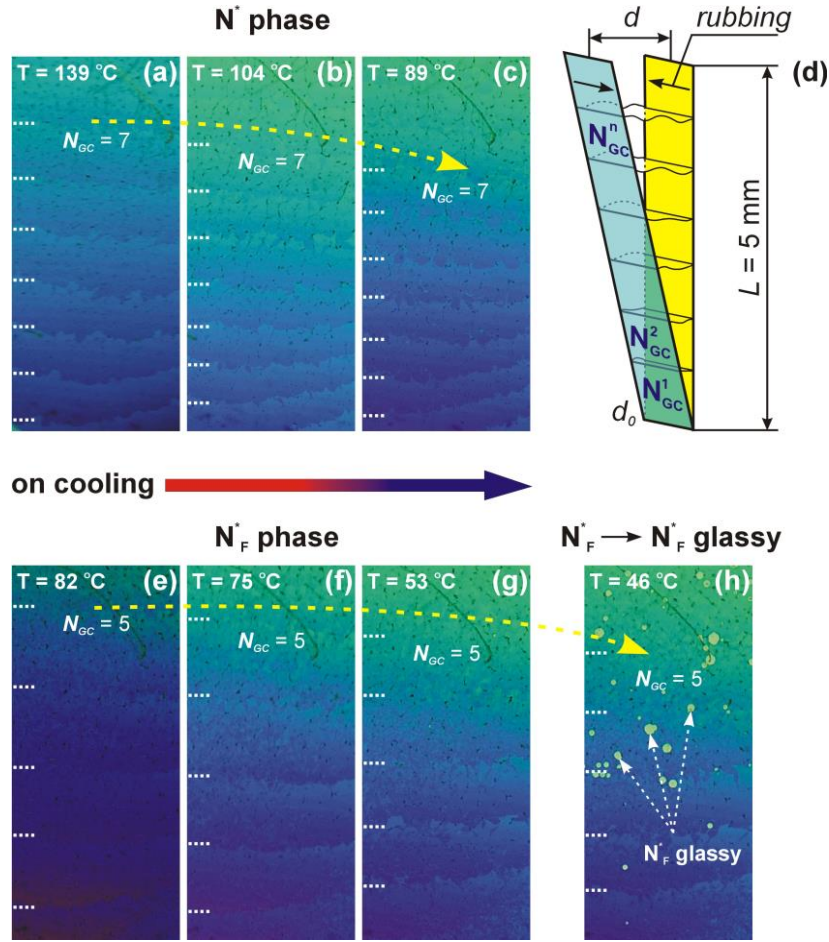

Figure 7. The sequential photographs of the wedge-like LC cell, filled by the $N^*_F$ mixture RM734:DIO (70:30) doped with 11.1 wt,% CB15 during the cooling process. (a)-(c) The increasing of $N_{GC}$ with decreasing temperature $T$ in $N^*$ phase of the studied $N^*_F$ mixture in wedge-like LC cell with thin end $d_0 = 2.7$ µm at $L \sim 0$ mm and thick end $d = 7.7$ µm at $L = 5$ mm.(d) (e)-(g) The increasing of $N_{GC}$ with decreasing temperature $T$ in $N^*_F$ phase. (g) Strongly cooled $N^*_F$ phase, accompanied by the appearance of areas characterized by $N^*_F$ glassy phase (indicated by white dashed arrows). The change of the GC line location ($N_{GC} = n$) during the cooling process is shown by yellow dashed arrows.

After $N^* \rightarrow N^*_F$ phase transition, which is accompanied by the substance flow (Figure 5 (b),(c)), the GC texture of the $N^*_F$ phase appears that also possesses RH helix (Figure 6 (e), (f)). The increasing of the $N_{GC}$ with decreasing temperature $T$ was found (Figure 7 (e)-(h)). The dependence of the calculated helical pitch $P$ of the $N^*_F$ phase on temperature $T$ is shown in Figure 8 (solid blue circles, curve 2). Consequently, in chiral ferronematic mixture containing CB15 the length of the helical pitch $P$ for both $N^*$ and

$N^*_F$ phase decreases with decreasing of the temperature $T$, in a similar way as recently reported in [36] for chiral ferronematic based on 6 wt.% CB15 dissolved in RM734. It should be noted that helical pitch $P$ can decrease in $N^*$ phase, while it increases in $N^*_F$ [36] or barely changed [19] with a decrease of temperature.

For comparison, the $P(T)$ dependence in the induced chiral nematic mixture (ChN), based on 11.4 wt.% of CB15 dissolved in E7 was measured in a similar way under the same experimental conditions. This dependence is also shown in Figure 8 (solid magenta spheres, curve 3).

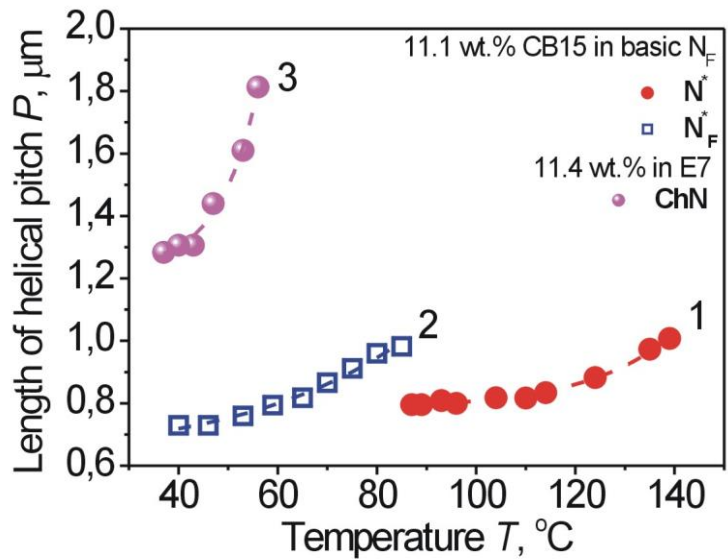

Figure 8. The temperature dependence of the length of helical pitch $P$ for both $N^*$ (solid red circles, curve 1) and $N^*_F$ (opened blue squares, curve 2) phase of the chiral ferroelectric mixture RM734:DIO (70:30) doped with 11.1 wt.% CB15 and chiral nematic mixture, based on 11.4 wt.% CB15 in nematic E7 (solid magenta spheres, curve 3). To calculate the length of the helical pitch $P$, the GC method of wedge-like LC cells was used. The thickness of thin end $d_0$ was in the range 1.7 - 3 µm and thick end was in the 6 – 7.6 µm at $L = 5$ mm.

### 3.3. Temperature-dependent optical transmission

The next step in our studies of ferroelectric nematic systems was related to measurements of optical transmission spectra as function of temperature. It is well known that optical transmission, *i.e.*, the fraction of incident light passing through a liquid crystal layer, is noticeably different in different phases. Also, peculiar features of optical transmission are often noted in the temperature ranges close to phase transitions due to various pre-transitional phenomena. This approach is especially useful for systems with nano-structuring, *e.g.*, for liquid crystal systems with dispersed nanoparticles [22-25] or supramolecular aggregation. [26] Recently, this method was used for systems forming $N_{tb}$ and heliconical (oblique cholesteric) phases [11].

A typical temperature dependence of optical transmission in the systems studied in the present work is shown in Figure 9. The transmittance values were taken at 700 nm to avoid interference with the absorption bands.

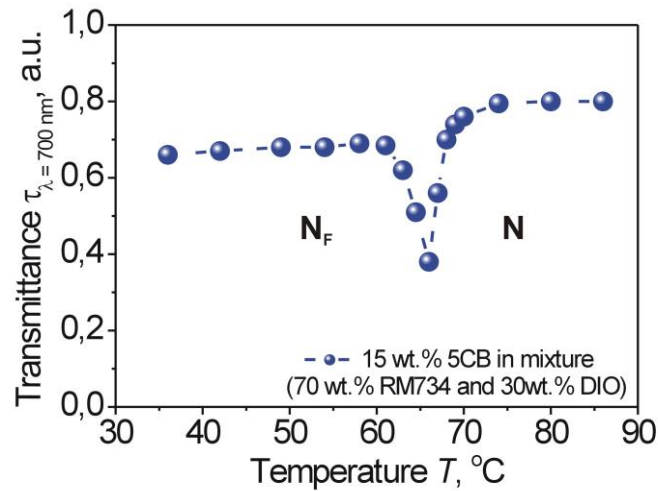

Figure 9. Optical transmittance for ferroelectric mixture $N_F$, based on 15 wt.% 5CB dissolved in the basic $N_F$ mixture, consisting of two ferroelectric nematics RM734 and DIO in ratio (70:30) at 700 nm as function of temperature. The thickness of the plane-parallel LC cell was 20 µm.

The transmittance in the ferroelectric nematic phase $N_F$ is lower as compared with the "ordinary" nematic phase N. This can be due to additional light scattering from the ferroelectric domain boundaries [5, 26]. Also, we see a sharp fall in transmission in the direct vicinity of the N → $N_F$ phase transition. To check such behavior, we made similar measurements in the kinetic mode, with the temperature continuously changing with time from a starting value in the N phase to the final value in the $N_F$ phase. The result obtained is shown in Figure 10.

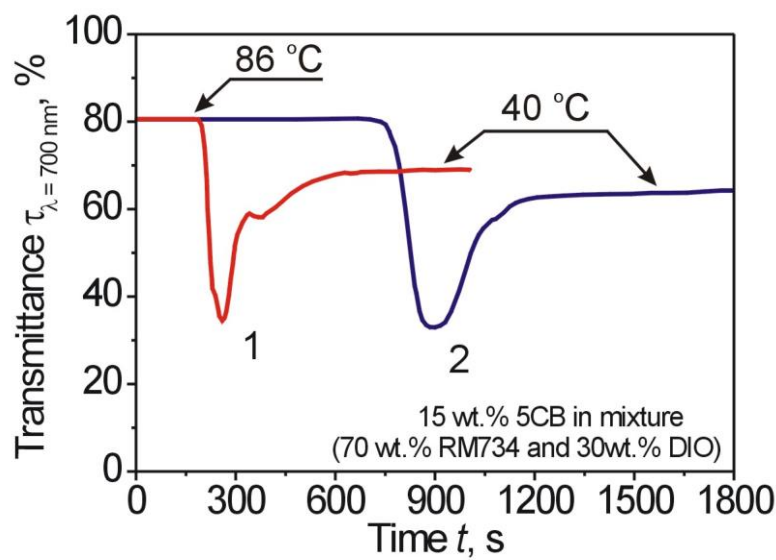

Figure 10. Variation of optical transmission with time upon cooling for the basic $N^*_F$ mixture, consisting of the two ferroelectric nematics RM734 and DIO in ratio (70:30), doped with 15 wt.% 5CB. The two plots correspond to different cooling rates.

The presence of the "intermediate" region with lower transmittance is most probably due to the processes of molecular reorientation during the phase transition, which can be not so straightforward, requiring further studies. However, one may assume that the main reason of such behavior may be the unstable dynamic processes (*e.g.,*dynamic substance flow [4]), such as observed in POM studies of the chiral ferronematic $N^*_F$ during its $N^* \to N^*_F$ transition (Figure 5).

**Conclusion**

The mixtures of ferronematic RM734 and DIO , as well as additional mesogenic and non-mesogenic components, were studied by DSC, POM and spectrophotometry. The basic ferronematic mixture, consisting of RM734 and DIO at the weight ratio of the components (70:30) was doped with both standard nematic 5CB and chiral dopant CB15. The introduction of these additives led to substantial lowering of the ferronematic temperature range in a similar way as with the same dopants in recent studies of $N_{tb}$ phases [11]. Addition of the chiral dopant CB15 led to induction of helical structure in both nematic (N) and ferronematic ($N_F$) phases, which were studied using the wedge-like LC cell to determine the screw sense of helical structure and the length of helical pitch by the Grandjean-Cano method. It was found that 5CB and CB15 were similar in their effects in the $N_F$ to N phase transition, while CB15 led to stronger decreasing of the isotropic transition temperature. During slow cooling process the Grandjean-Cano texture of chiral nematic phase does not disappear in the chiral ferronematic phase, while fast cooling gave rise to formation of undulation structures. It was found that the decreasing of temperature leads to decreasing of length of helical pitch in both chiral nematic and chiral ferronematic phase, the helical pitch decreased upon cooling, with this dependence being similar to that observed in "ordinary" nematic phase. Optical transmission vs. temperature measurements showed lower transmittance in the $N_F$ phase, probably due to scattering on the domain boundaries and unstable dynamic process caused by the substance flow. In the vicinity of the $N_F$ - N transition, a noticeable range of anomalously low transmission was noted both on heating and cooling, suggesting a complex character of molecular reorientation features.

Acknowledgements

The research was supported by the National Academy of Sciences of Ukraine within projects № 0122U002636, 3/2022-B and B#197. The authors thank W. Becker (Merck, Darmstadt, Germany) for his generous gift of nematic liquid crystal E7 and field service specialist V. M. Danylyuk (Dish LLC, USA) for his gift of some Laboratory equipments.